# *In-situ* observations of resident space objects with the CHEOPS space telescope

Nicolas Billot[a,*], Stephan Hellmich[b], Willy Benz[c,d], Andrea Fortier[c,d], David Ehrenreich[a,e], Christopher Broeg[c,d], Alexis Heitzmann[a], Anja Bekkelien[a], Alexis Brandeker[f], Yann Alibert[d,c], Roi Alonso[g,h], Tamas Bárczy[i], David Barrado Navascues[j], Susana C.C. Barros[k,l], Wolfgang Baumjohann[m], Federico Biondi[n,o], Luca Borsato[o], Andrew Collier Cameron[p], Carlos Corral van Damme[q], Alexandre C.M. Correia[r], Szilard Csizmadia[s], Patricio E. Cubillos[m,t], Melvyn B. Davies[u], Magali Deleuil[v], Adrien Deline[a], Olivier D.S. Demangeon[k,l], Brice-Olivier Demory[d,c], Aliz Derekas[w], Billy Edwards[x], Jo Ann Egger[c], Anders Erikson[s], Luca Fossati[m], Malcolm Fridlund[y,z], Davide Gandolfi[aa], Kosmas Gazeas[bb], Michaël Gillon[cc], Manuel Güdel[dd], Maximilian N. Günther[q], Ch. Helling[m,ee], Kate G. Isaak[q], Laszlo L. Kiss[ff,gg], Judith Korth[hh], Kristine W.F. Lam[s], Jacques Laskar[ii], Alain Lecavelier des Etangs[jj], Monika Lendl[a], Demetrio Magrin[o], Pierre F.L. Maxted[kk], Marko Mecina[ll], Bruno Merín[mm], Christoph Mordasini[c,d], Valerio Nascimbeni[o], Göran Olofsson[f], Roland Ottensamer[dd], Isabella Pagano[nn], Enric Pallé[g,h], Gisbert Peter[oo], Daniele Piazza[c], Giampaolo Piotto[o,pp], Don Pollacco[qq], Didier Queloz[rr,ss], Roberto Ragazzoni[o,pp], Nicola Rando[q], Heike Rauer[s,tt], Ignasi Ribas[uu,vv], Martin Rieder[ww,d], Nuno C. Santos[k,l], Gaetano Scandariato[nn], Damien Ségransan[a], Attila E. Simon[c,d], Alexis M.S. Smith[s], Sérgio G. Sousa[k], Manu Stalport[xx,cc], Sophia Sulis[v], Gyula M. Szabó[w,yy], Stéphane Udry[a], Bernd Ulmer[oo], Solène Ulmer-Moll[a], Valérie Van Grootel[xx], Julia Venturini[a], Eva Villaver[g,h], Nicholas A. Walton[zz], Thomas G. Wilson[qq]

[a] *Observatoire astronomique de l'Université de Genève, Chemin Pegasi 51, Versoix 1290, Switzerland*
[b] *EPFL Laboratory of Astrophysics (LASTRO), Chemin Pegasi 51, Versoix 1290, Switzerland*
[c] *Weltraumforschung und Planetologie, Physikalisches Institut, University of Bern, Gesellschaftsstrasse 6, Bern 3012, Switzerland*
[d] *Center for Space and Habitability, University of Bern, Gesellschaftsstrasse 6, Bern 3012, Switzerland*
[e] *Centre Vie dans l'Univers, Faculté des Sciences, Université de Genève, Quai Ernest-Ansermet 30, 1211 Genève 4, Switzerland*
[f] *Department of Astronomy, Stockholm University, AlbaNova University Center, Stockholm 10691, Sweden*
[g] *Instituto de Astrofísica de Canarias, Vía Láctea s/n, Tenerife, La Laguna 38200, Spain*
[h] *Departamento de Astrofísica, Universidad de La Laguna, Astrofísico Francisco Sanchez s/n, Tenerife, La Laguna 38206, Spain*
[i] *Admatis, 5. Kandó Kálmán Street, Miskolc 3534, Hungary*
[j] *Depto. de Astrofísica, Centro de Astrobiología (CSIC-INTA), ESAC campus, 28692 Villanueva de la Cañada (Madrid), Spain*
[k] *Instituto de Astrofisica e Ciencias do Espaco, Universidade do Porto, CAUP, Rua das Estrelas, Porto 4150-762, Portugal*
[l] *Departamento de Fisica e Astronomia, Faculdade de Ciencias, Universidade do Porto, Rua do Campo Alegre, Porto 4169-007, Portugal*
[m] *Space Research Institute, Austrian Academy of Sciences, Schmiedlstrasse 6, Graz A-8042, Austria*
[n] *Max Planck Institute for Extraterrestrial Physics, Gießenbachstraße 1, Garching bei München 85748, Germany*
[o] *INAF, Osservatorio Astronomico di Padova, Vicolo dell'Osservatorio 5, Padova 35122, Italy*
[p] *Centre for Exoplanet Science, SUPA School of Physics and Astronomy, University of St Andrews, North Haugh, St Andrews KY16 9SS, UK*
[q] *European Space Agency (ESA), European Space Research and Technology Centre (ESTEC), Keplerlaan 1, AZ Noordwijk 2201, the Netherlands*
[r] *CFisUC, Departamento de Física, Universidade de Coimbra, Coimbra 3004-516, Portugal*
[s] *Institute of Planetary Research, German Aerospace Center (DLR), Rutherfordstrasse 2, Berlin 12489, Germany*
[t] *INAF, Osservatorio Astrofisico di Torino, Via Osservatorio, 20, Pino Torinese To I-10025, Italy*
[u] *Centre for Mathematical Sciences, Lund University, Box 118, Lund 221 00, Sweden*
[v] *Aix Marseille Univ, CNRS, CNES, LAM, 38 rue Frédéric Joliot-Curie, Marseille 13388, France*
[w] *ELTE Gothard Astrophysical Observatory, 9700 Szombathely, Szent Imre h. u. 112, Hungary*

* Corresponding author.
*E-mail address:* nicolas.billot@unige.ch (N. Billot).

[x] SRON Netherlands Institute for Space Research, Niels Bohrweg 4, CA Leiden 2333, the Netherlands
[y] Leiden Observatory, University of Leiden, PO Box 9513, RA Leiden 2300, the Netherlands
[z] Department of Space, Earth and Environment, Onsala Space Observatory, Chalmers University of Technology, Onsala 439 92, Sweden
[aa] Dipartimento di Fisica, Università degli Studi di Torino, via Pietro Giuria 1, Torino I-10125, Italy
[bb] Department of Physics, University Campus, Zografos GR-157 84, Athens, Greece
[cc] Astrobiology Research Unit, Université de Liège, National and Kapodistrian University of Athens, Allée du 6 Août 19C, Liège B-4000, Belgium
[dd] Department of Astrophysics, University of Vienna, Türkenschanzstrasse 17, Vienna 1180, Austria
[ee] Institute for Theoretical Physics and Computational Physics, Graz University of Technology, Petersgasse 16, Graz 8010, Austria
[ff] Konkoly Observatory, Research Centre for Astronomy and Earth Sciences, Konkoly Thege Miklós út 15-17, Budapest 1121, Hungary
[gg] Institute of Physics, ELTE Eötvös Loránd University, Pázmány Péter sétány 1/A, Budapest 1117, Hungary
[hh] Lund Observatory, Division of Astrophysics, Department of Physics, Lund University, Box 118, Lund 22100, Sweden
[ii] IMCCE, UMR8028 CNRS, Observatoire de Paris, PSL University, Sorbonne University, 77 av. Denfert-Rochereau, Paris 75014, France
[jj] UMR7095 CNRS, Institut d'astrophysique de Paris, Université Pierre & Marie Curie, 98bis blvd. Arago, Paris 75014, France
[kk] Astrophysics Group, Lennard Jones Building, Keele University, Staffordshire ST5 5BG, UK
[ll] Department of Astrophysics, Tuerkenschanzstr. 17, Vienna 1180, Austria
[mm] European Space Agency, ESA - European Space Astronomy Centre, Camino Bajo del Castillo s/n, 28692 Villanueva de la Cañada, Madrid, Spain
[nn] INAF, Osservatorio Astrofisico di Catania, Via S. Sofia 78, Catania 95123, Italy
[oo] Institute of Optical Sensor Systems, German Aerospace Center (DLR), Rutherfordstrasse 2, Berlin 12489, Germany
[pp] Dipartimento di Fisica e Astronomia "Galileo Galilei", Università degli Studi di Padova, Vicolo dell'Osservatorio 3, Padova 35122, Italy
[qq] Department of Physics, University of Warwick, Gibbet Hill Road, Coventry CV4 7AL, UK
[rr] ETH Zurich, Department of Physics, Wolfgang-Pauli-Strasse 2, Zurich CH-8093, Switzerland
[ss] Cavendish Laboratory, JJ Thomson Avenue, Cambridge CB3 0HE, UK
[tt] Institut fuer Geologische Wissenschaften, Freie Universitaet Berlin, Maltheserstrasse 74-100, Berlin 12249, Germany
[uu] Institut de Ciencies de l'Espai (ICE, CSIC), Campus UAB, Can Magrans s/n, Bellaterra 08193, Spain
[vv] Institut d'Estudis Espacials de Catalunya (IEEC), 08860 Castelldefels (Barcelona), Spain
[ww] Weltraumforschung und Planetologie, Physikalisches Institut, University of Bern, Sidlerstrasse 5, Bern 3012, Switzerland
[xx] Space sciences, Technologies and Astrophysics Research (STAR) Institute, Université de Liège, Allée du 6 Août 19C, Liège 4000, Belgium
[yy] HUN-REN-ELTE Exoplanet Research Group, Szent Imre h. u. 112., Szombathely H-9700, Hungary
[zz] Institute of Astronomy, University of Cambridge, Madingley Road, Cambridge CB3 0HA, UK

ABSTRACT

The CHaracterising ExOPlanet Satellite (CHEOPS) is a partnership between the European Space Agency and Switzerland with important contributions by 10 additional ESA member States. It is the first S-class mission in the ESA Science Programme. CHEOPS has been flying on a Sun-synchronous low Earth orbit since December 2019, collecting millions of short-exposure images in the visible domain to study exoplanet properties.

A small yet increasing fraction of CHEOPS images show linear trails caused by resident space objects crossing the instrument field of view. CHEOPS' orbit is indeed particularly favourable to serendipitously detect objects in its vicinity as the spacecraft rarely enters the Earth's shadow, sits at an altitude of 700 km, and observes with moderate phase angles relative to the Sun. This observing configuration is quite powerful, and it is complementary to optical observations from the ground.

To characterize the population of satellites and orbital debris observed by CHEOPS, all and every science images acquired over the past 3 years have been scanned with a Hough transform algorithm to identify the characteristic linear features that these objects cause on the images. Thousands of trails have been detected. This statistically significant sample shows interesting trends and features such as an increased occurrence rate over the past years as well as the fingerprint of the Starlink constellation. The cross-matching of individual trails with catalogued objects is underway as we aim to measure their distance at the time of observation and deduce the apparent magnitude of the detected objects.

As space agencies and private companies are developing new space-based surveillance and tracking activities to catalogue and characterize the distribution of small debris, the CHEOPS experience is timely and relevant. With the first CHEOPS mission extension currently running until the end of 2026, and a possible second extension until the end of 2029, the longer time coverage will make our dataset even more valuable to the community, especially for characterizing objects with recurrent crossings.

## 1. The characterising exoplanet satellite (CHEOPS) mission

The European Space Agency (ESA) and the CHEOPS Consortium have partnered up to build and operate the first S-class mission of the ESA Science program. The CHEOPS Consortium is composed of 11 European countries, led by the University of Bern, Switzerland. CHEOPS was launched from the European Space Port in Kourou, French Guyana, by a Soyouz rocket on 18th December 2019. CHEOPS was successfully injected into its nominal Sun-synchronous orbit at 700 km altitude with a Local Time of Ascending Node of 6 a.m.

CHEOPS was designed to observe and characterize exoplanets with an exquisite precision of 10 parts-per-million (ppm) on targets brighter than GAIA magnitude 9. CHEOPS is equipped with a telescope of effective aperture 30 cm, operating in the visible range between 400 and 800 nm.

The camera is equipped with a 1024x1024 CCD, each pixel projecting to 1 arcsecond on the sky. Science images are cropped onboard to spare bandwidth and only 200x200 sub-arrays are downlinked to the ground. Exposure times vary between 1 ms and 60 s, depending on the brightness of the targeted star, with a typical value of about 20 s and a 25 ms deadtime between consecutive exposures. The absolute timing accuracy of an exposure is typically 60 ms (always below 200 ms) for the start time, and better than 0.12 ms for the duration of the exposure.

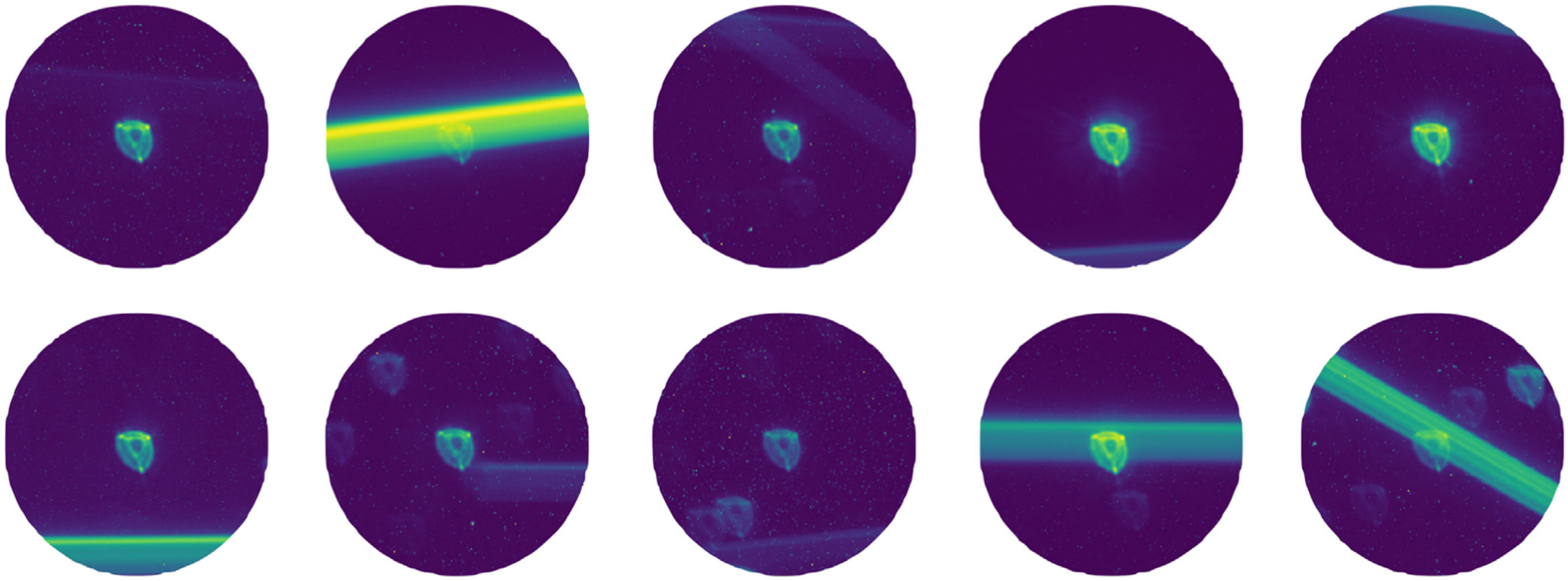

**Fig. 1.** Gallery of satellite/debris trails observed in CHEOPS science images showing bright, faint, grazing, partial and multiple trails.

CHEOPS' observing mode consists in staring at one star at a time, for periods typically ranging from hours to days, to acquire a series of short-exposure images from which the target flux is measured precisely. Resulting light curves are exploited to detect variations that are primarily caused by planet(s) orbiting the host star. Planetary parameters such as radius, refined ephemeris, atmospheric properties can then be extracted from such light curves.

A complete description of the CHEOPS mission, including its platform, payload, ground-segment organization as well as its science program is given in [1] and [2].

## 2. Resident space objects crossing CHEOPS' field-of-view

In early January 2020, only a few hours after having opened the telescope tube cover, still during the commissioning phase, we observed for the first time a bright object that had crossed the field-of-view of the instrument at high velocity leaving behind a trail across the entire image. Since then, we have observed thousands of those, several per day, affecting the science images of CHEOPS. Fig. 1 shows a gallery of such trails that we have collected.

The feature at the center of each image in Fig. 1 is the target star that was being observed at the time of the object crossing. As the CHEOPS instrument has been intentionally defocused by design to enhance its sensitivity, the Point Spread Function (PSF) spreads over 16 arcsec in radius and it has a complex structure, with a roundish halo and three bright spots due to the 3-legged structure that holds the secondary mirror. Details of the CHEOPS PSF and the overall instrument performances can be found in [2].

## 3. Automatic detection algorithm

In the present analysis, we are considering all and every CHEOPS science observation carried out over a period of over 3 years covering March 2020 until May 2023. This represents over 1.25 million images collected over this period as part of the science program. These images are 200 arcsec across. They have been acquired with exposure times ranging from 1 ms for the brightest target stars to 60 s for the faintest ones, with a typical value around 30 s.

To cope with the large number of images, we have set up an automatic pipeline that scans each individual image and searches for linear features characteristic of satellite/debris trails.

Some pre-processing steps are required to ease and make the trail detection process more robust:

- Apply a median filter to the image to remove hot pixels
- Normalise the image to increase its contrast
- Apply a Canny edge detection to further increase the contrast. This step requires the fine tuning of thresholds to detect appropriate intensity gradients in the CHEOPS images and obtain the expected results.

We finally execute a Hough transform on the pre-processed images to detect and characterize trails (extract position angle and location on the image).

A dedicated completeness analysis of the trail detection process is underway. We have nevertheless already identified the root cause of most false positive detections:

- Excessive straylight from the Earth dayside may generate structures in the image background, e.g. linear ripples. Those images are excluded from the analysis.
- Edge-on cosmic ray hits lead to linear features in images. Those features can easily be filtered out as they are about 1-pixel wide, significantly thinner than a trail caused by a crossing debris.
- Crowded stellar fields, in combination with the spatial features of the CHEOPS PSF, may lead to false detections. These erroneous trail detections occur solely at 45 and 135° (alignment of structural features of the PSF), they are thus straightforward to filter out.
- Rare cases of saturation lead to vertical features that can be mis-identified for debris trails. Filtering out trails with position angles equal to zero solves this issue.
- CHEOPS' camera being shutterless, images containing very bright stars suffer from vertical smearing features. We can remedy this issue as above, filtering out images with 0° position angles.

The last caveat we have identified in our detection process concerns multiple trails. Those are currently considered as single trails with erroneous position/angle. Note that those represent only a small fraction of all trails seen in CHEOPS' field-of-view, estimated to a few percent at most, thus not biasing the statistical analysis presented below.

Out of the 1.25 million images scanned with our pipeline, and after filtering out false positives, we have identified a total of 3200 images that contain a trail due to a satellite or space debris crossing the field-of-view. This represents a mere 0.26% of the whole dataset, and this confirms that the CHEOPS Science program is not currently at risk due to these unwanted trails.

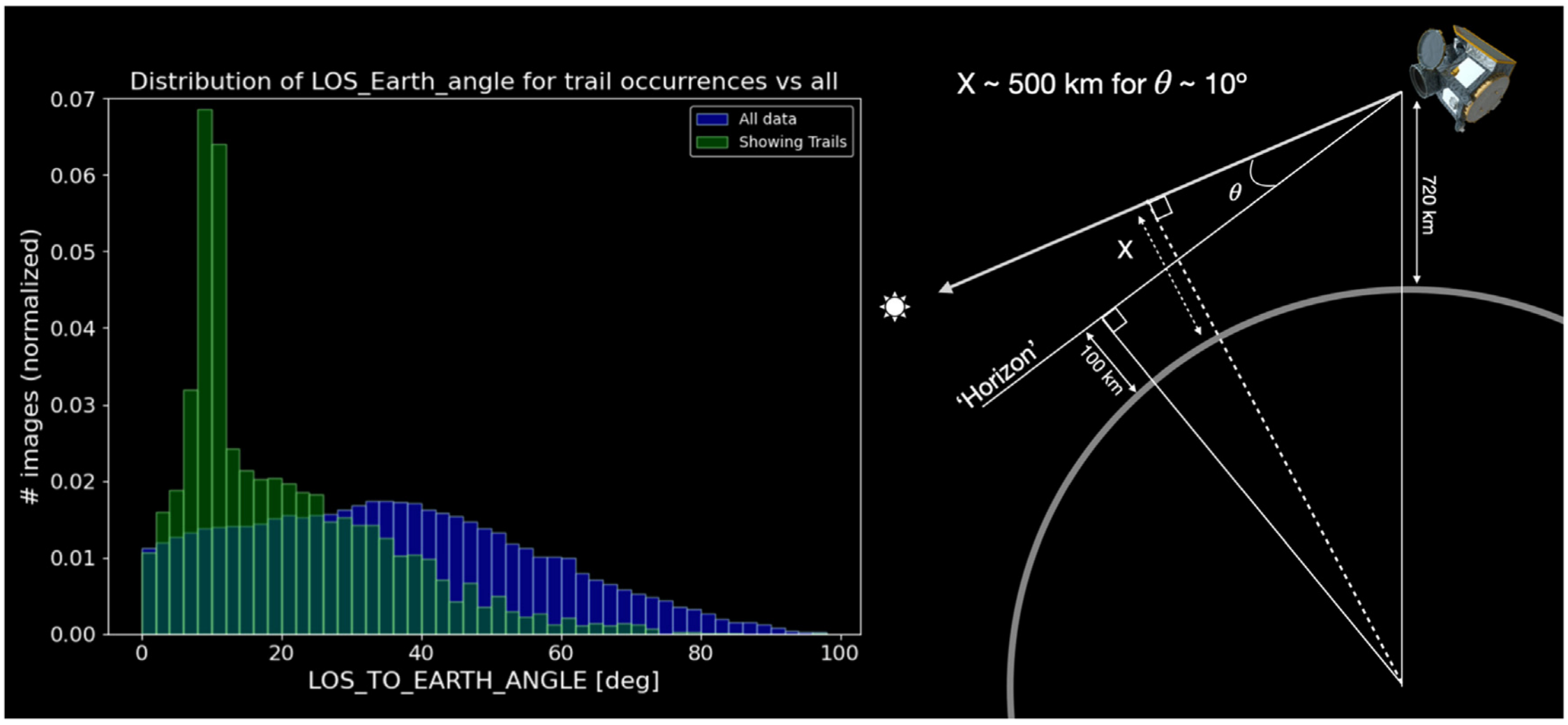

**Fig. 2.** Distribution of angular distances between line-of-sight (LOS) and Earth horizon, as depicted in right diagram.

This relatively low value, compared to the few percents quoted in e.g. [3] for the Hubble Space Telescope, are mainly due to the very small field-of-view of CHEOPS as well as the relatively short exposure times compared to those typically used with Hubble. Other elements, such as the spacecraft orbit, also impact the occurrence rate of trails in images.

## 4. Dataset analysis

With thousands of trail detections, we have a large enough sample to explore possible trends on these crossing events. In the following sub-sections, we use the metadata available from CHEOPS images, like the line-of-sight (LOS) relative to the Earth horizon, to enrich our dataset and highlight statistically significant features.

### *4.1. Shell of debris in low earth orbit*

Along its revolution around the Earth, CHEOPS' line-of-sight remains constant in the **International Celestial Reference Frame** (**ICRF**) for the duration of an observation, typically for a few hours/days. CHEOPS therefore probes various angles above the Earth horizon during the course of an observation. Fig. 2 shows the distribution of the angular distance between the line-of-sight and the Earth horizon.

The blue histogram shows the distribution for the entire dataset of 1.25 million images. The green histogram shows the distribution exclusively for images affected by trails. Both distributions are strikingly different, which means that trails occur preferentially at some specific angles above the horizon. In particular, the green histogram of Fig. 2 shows a peak at around 10° above the horizon which denotes an excess detection rate at a specific '*angular altitude*' above the horizon. Depending on where along the line-of-sight the object is crossing, as depicted in Fig. 2, an elevation of 10° corresponds to an object orbiting at 500 km or above, typically between 500 and 700 km. This peak in the green histogram traces a high-density shell of satellites and debris, corresponding to the lower part of the so-called Low Earth Orbit, that has been nicely probed by the CHEOPS line-of-sight over the past 3 years.

Ground-based observations of LEO objects show a similar peak just above the horizon when the line-of-sight probes the largest volume of the LEO shell, as demonstrated in [4]. As CHEOPS sits in LEO, it offers a complementary view of the situation.

### *4.2. Higher detection rates for small phase angles*

CHEOPS' orbit is Sun-synchronous such that the Sun always shines on the backside of the spacecraft to minimize straylight contamination (and power up solar arrays). Fig. 3 shows the distribution of the angular distance between the line-of-sight and the Sun direction, which defines the angle $\theta$ as depicted in the right panel of the figure. Allowed values range from 120 to 180°, meaning that we only observe within 60° of the anti-Sun direction.

The blue histogram shows the distribution for the entire dataset of 1.25 million images. The green histogram shows the distribution exclusively for images affected by trails. Both distributions are different, with an excess detection rate at low phase angles, i.e. when the crossing object is illuminated from a direction close to the line-of-sight. We interpret this feature as meaning that crossing objects look globally brighter when the Sun shines on them from straight behind the observer. This is consistent with most phase functions that describe the dependence of scattered light versus the incident radiation phase angle.

### *4.3. Starlink imprint in geographic coordinates*

We now consider the geographic position of CHEOPS at the time of observation to search for trends and features. Fig. 4 shows the distribution of latitudes at which CHEOPS was located when it was observing.

The blue histogram shows the distribution for the entire dataset of 1.25 million images. Note that, with an orbital inclination of 98.7°, CHEOPS always remains below a latitude of 81.3°. The asymmetry close to the poles is also due to the inclination of CHEOPS' orbit as the spacecraft spends on average more time on the day side of the Earth when close to the South pole than close to the North pole. As CHEOPS is very sensitive to straylight, images acquired close to the South pole are often affected by straylight and therefore rejected on-board, thus leading to a deficit of images at very low latitudes. In addition, the dip in the blue distribution between latitudes 0 and -50° is due to the South Atlantic Anomaly (SAA). When CHEOPS crosses the SAA, data are not recorded on-board because of a too high level of cosmic ray hits on the detector.

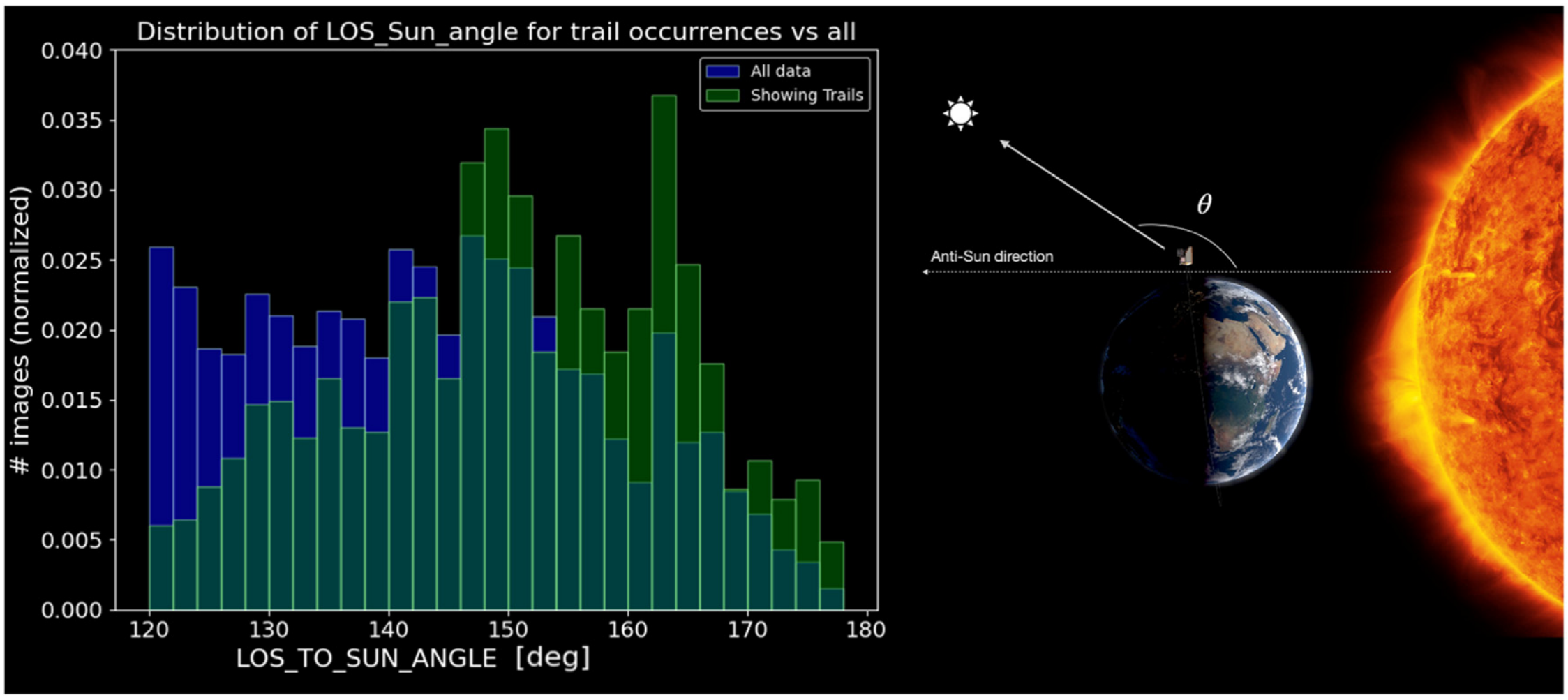

**Fig. 3.** Distribution of angular distances between line-of-sight (LOS) and Sun direction.

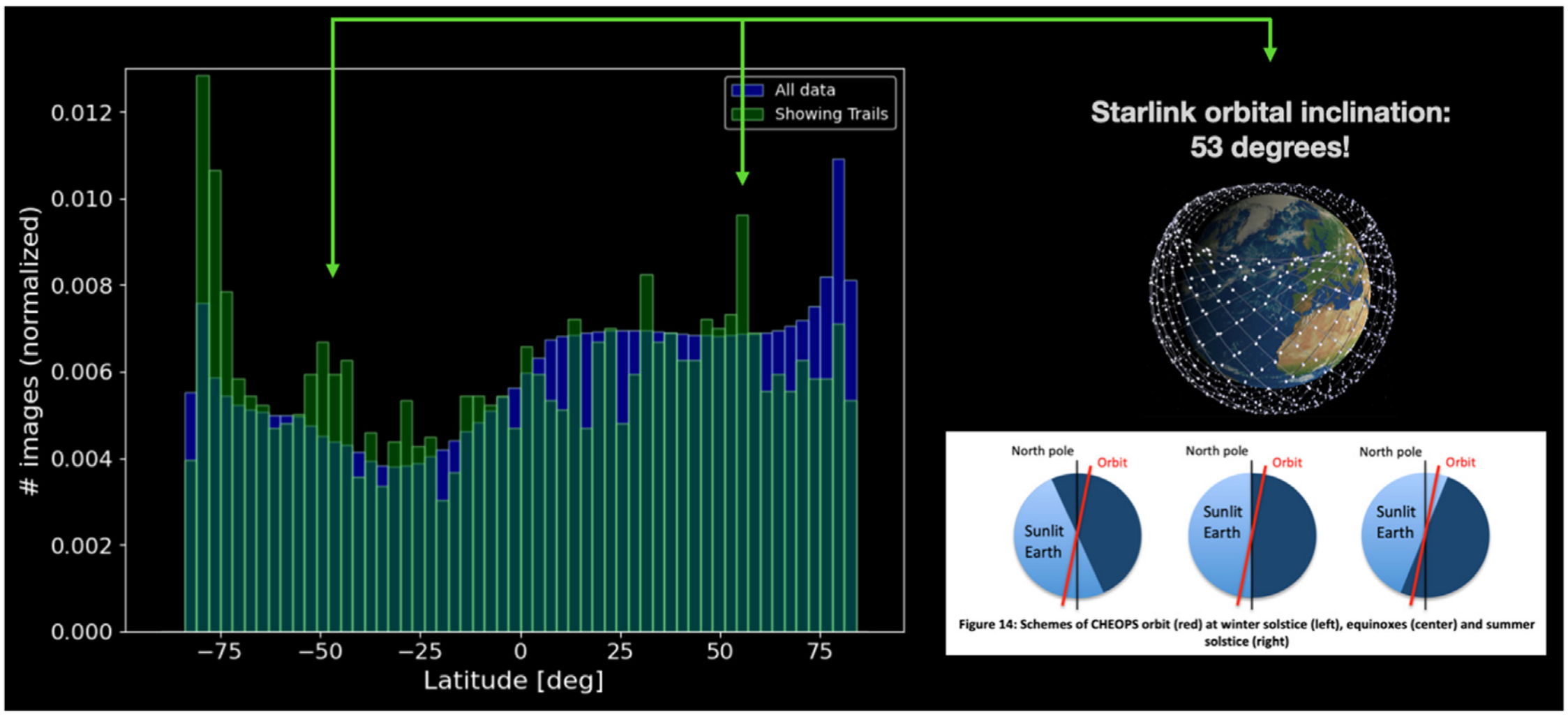

**Fig. 4.** Distribution of latitudes at time of observations.

The green histogram shows the distribution of the same quantity but exclusively for images affected by trails. Both distributions are again quite different. The asymmetry between South and North poles, also caused by CHEOPS' orbital inclination, goes in the opposite direction now. Close to the North pole, CHEOPS spends a fraction of its time in the shadow of the Earth during winter months, in the so-called eclipse season. During this period, a fraction of the objects crossing the field-of-view of CHEOPS will go unnoticed as they would also be in the shadow of the Earth during observations. This explains the deficit of trail detections close to the North pole. Conversely, when CHEOPS is close to the South pole at around -80° latitude, objects crossing the field-of-view spend comparatively more time illuminated by the Sun, over the course of the year (see bottom right panel of Fig. 4), thus increasing the total number of detections at those latitudes.

The green distribution also shows a couple of peaks at around -50 and +50° latitude. Coincidentally a large fraction of the Starlink constellation have an inclination of 53°, which means that Starlink satellites spends comparatively more time at latitudes ± 53° than at other latitudes thus increasing the likelihood of being observed by CHEOPS at those latitudes. We therefore interpret the couple of features at ± 53° latitude as being the imprint of Starlink.

### 4.4. Increased trail occurrence rate

With the injection of thousands of Starlink satellites in low Earth orbit over the past couple of years, we expect to see an increase in the number of trails detected with CHEOPS. This is indeed what is observed and presented in Fig. 5 where we plot the bi-weekly number count of trail detections.

The histogram shows strong modulations due to seasonal effects, with a severe deficit of detections during winter months. This is related to CHEOPS entering the eclipse season, as mentioned in Section 4.3, where the spacecraft spends a fraction of its time in the Earth's shadow. During these winter months, many objects crossing the CHEOPS field-of-view will also be in the shadow of the Earth thus making them generally invisible or too faint[1] to be detected in the images, and consequently reducing the number of detections. Summer months suffer from a similar effect, but to a lesser extent.

Independently of these seasonal effects, the overall trend seen in Fig. 5 is indeed an increase in the detection rate. This trend becomes even clearer when considering specific 'angular altitudes' above the Earth horizon.

[1] Moonlight, atmospheric airglow or aurorae are very faint light sources.

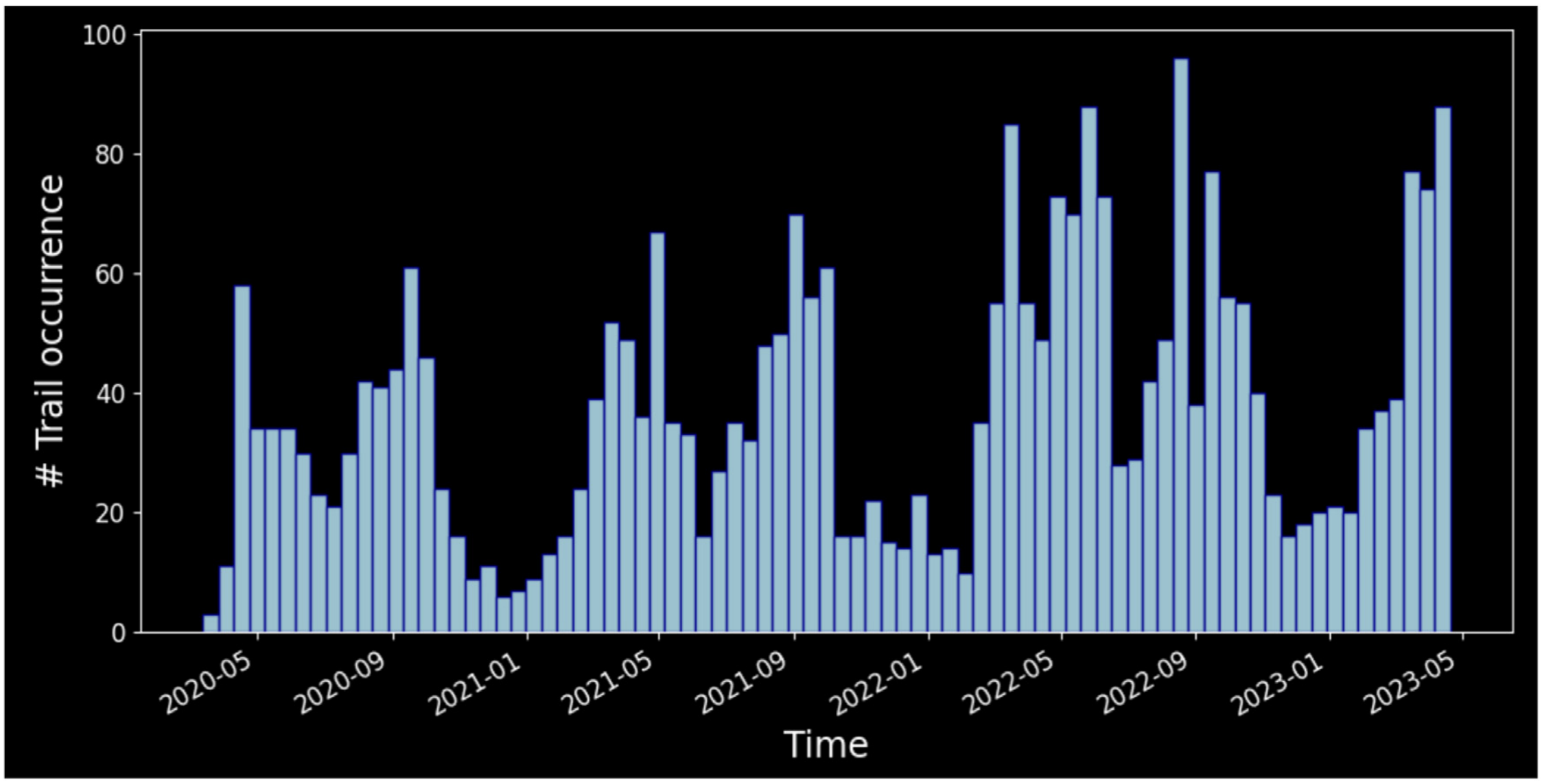

**Fig. 5.** Trail detection number count in bi-weekly bins.

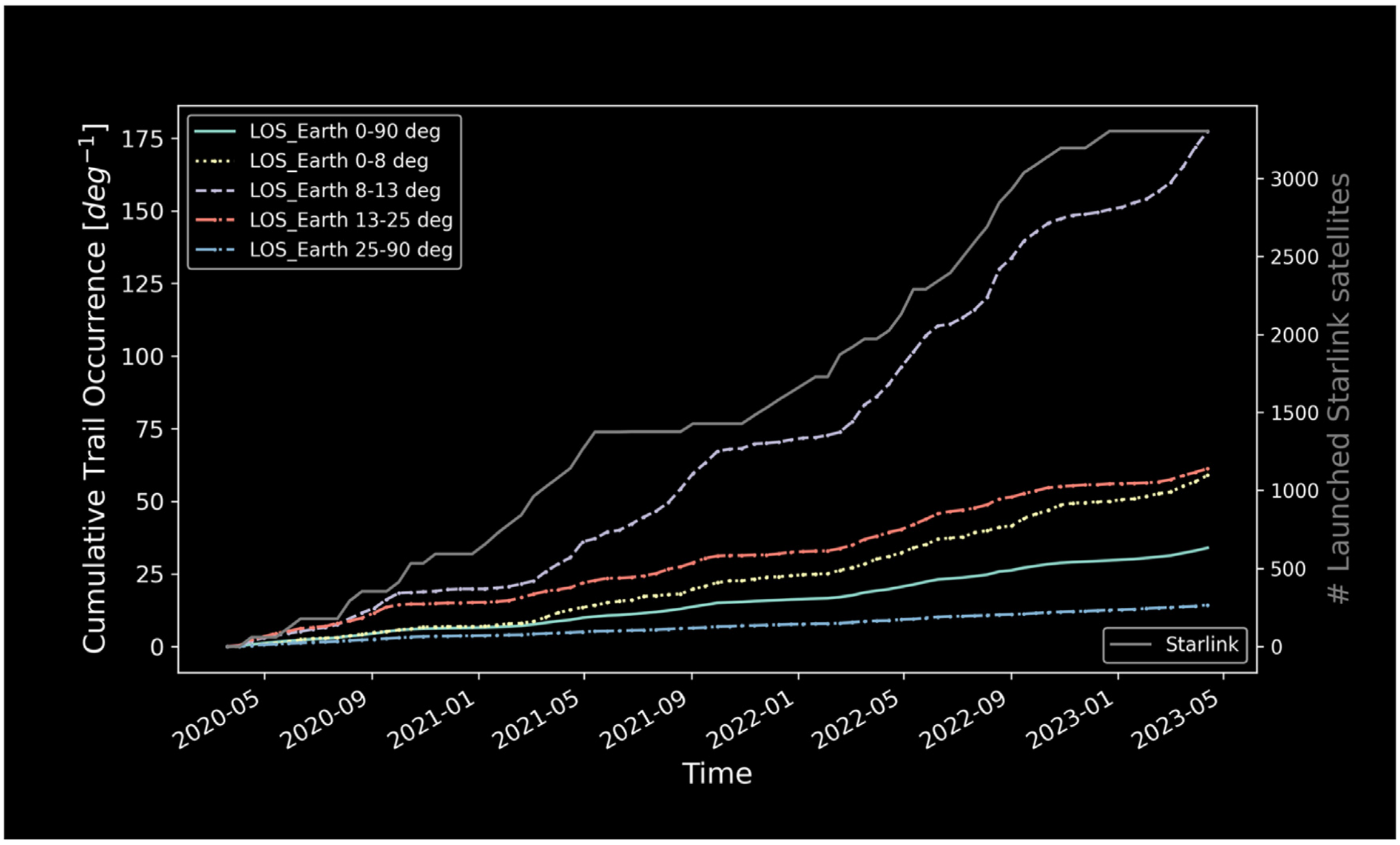

**Fig. 6.** Cumulative distribution of trail occurrences for various angular distances between line-of-sight and Earth horizon, along with the number of Starlink satellites injected into low Earth orbit over the same period.

Fig. 6 presents the cumulative distribution of trail occurrences for various bins of angular distances between the line-of-sight and the Earth horizon, as described in Section 4.1. The distribution is normalized by the bin size to represent a proxy of the debris/satellite density in each bin.

When considering only trails detected far above the horizon, e.g. between 25 and 90°, Fig. 6 shows that the cumulative distribution is nearly linear. This means that the number of detections, per degree above the horizon, remains relatively constant, i.e. the population of objects detected in this zone has changed very little, if at all, over the monitored period.

Conversely, the densest zone probed between 8 and 13° above the horizon, corresponding to the peak in Fig. 2, shows the strongest increase in the number of trail detections. As this zone traces the population of resident space objects in low Earth orbit, we interpret this steep rise as a significant increase in the number of objects now populating this zone. This is consistent with the fact that Starlink has launched thousands of satellites into low Earth orbit over the past couple of years, as shown in Fig. 6.

## 5. Object identification

For a detailed analysis of each individual detection, the observational circumstances, such as observer-target distance, phase angle and angular velocity are required. To obtain this information, we linked each observed trail to the catalog of known space objects provided by the US 18th Space Defense Squadron, published on the space-track website[2]. Orbital information on space objects is usually provided by Two-Line Element sets (TLEs), that correspond to the mean keplerian orbital elements at a given epoch. This results in the accuracy of positions computed from propagating the TLEs to decrease with increasing distance from the TLE epoch.

[2] https://www.space-track.org/.

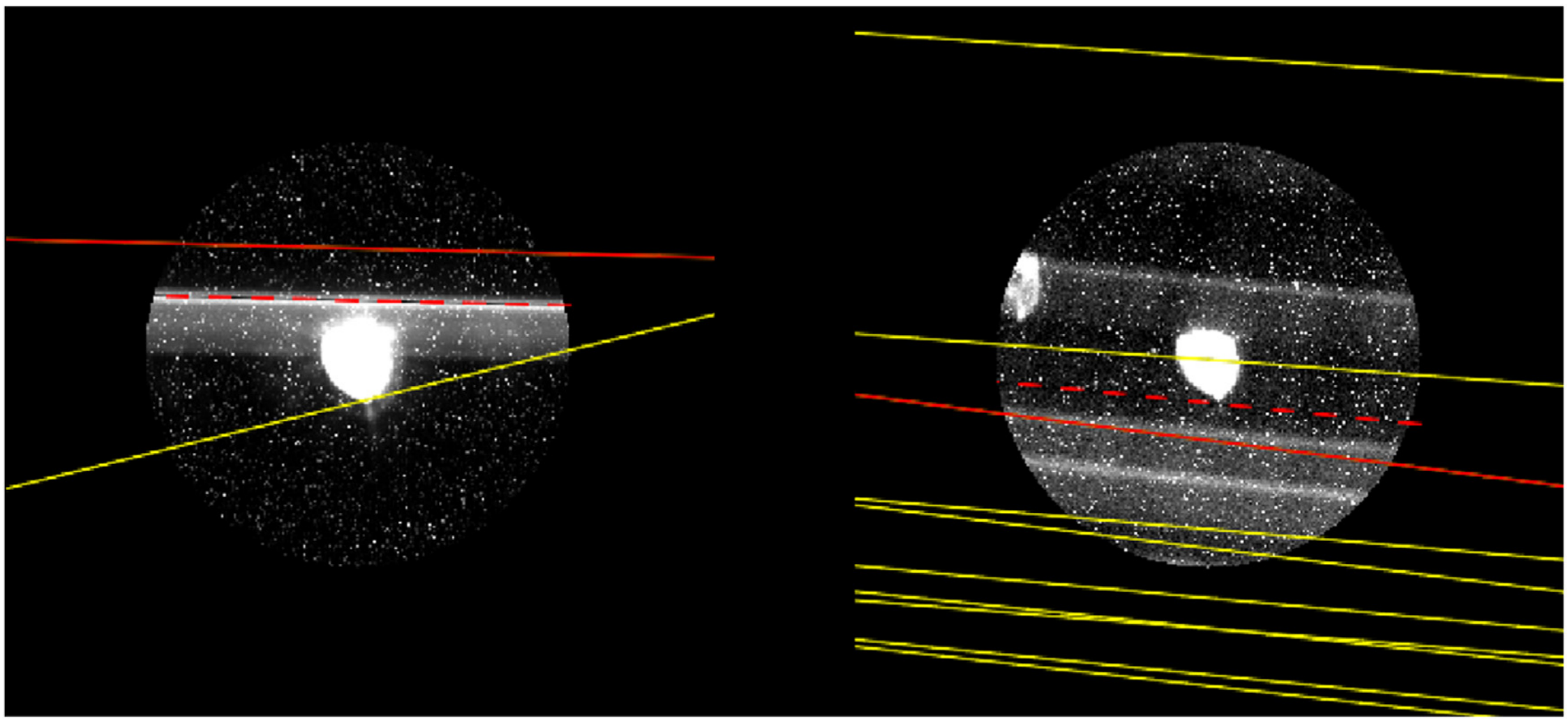

**Fig. 7.** Trail detections (dashed red lines), projections of cataloged objects that cross the field of view (solid lines) and the assigned known object (solid red line). Left: correct identification with multiple candidates. Right: Impossible identification due to too many candidates.

In order to minimize propagation errors in the identification process, for each CHEOPS image that contains a trail, we selected the TLE of CHEOPS that was closest to the observation time to compute the observer location, and for every cataloged space object, we selected its TLE closest to the observation time from an interval of 2 days around the observation epoch. This resulted in about 18,000 to 25,000 TLEs for each observation. Using the precise observation time and the astrometric solution stored along with the CHEOPS data, we were able to compute which of those objects were crossing the observed field during the exposure and projected their track into the image. This projection is of limited precision due to two reasons. Firstly, because the observer location and the positions of the cataloged space objects were computed from TLEs which are inherently imprecise. The uncertainty on positions, computed by propagating the orbit, grows with increasing time to/since the TLE epoch. This leads to the offset between observed trail and the projected TLEs. Secondly because CHEOPS is nearly nadir-locked, i.e. slowly rotates around the optical axis in order keep its radiators always away from Earth. The astrometric solution that is provided along with the data and used for the TLE projection corresponds to the orientation of the spacecraft at the start of the exposure. This means that the rotation of the spacecraft throughout the exposure is not accounted for when projecting the TLEs and results in the position angle of the projected TLE may be slightly different from that of the detected trail. However, by defining margins for the identification, it was still possible to reliably assign the majority of the trails to cataloged objects (left panel in Fig. 7). The only two occasions where we identified ambiguities were for starlink satellites just after their deployment (starlink trains). In this case, many satellites are crossing the field of view in the same direction and because the trail detection algorithm can not detect multiple trails in a single image, a unique assignment is not possible (right panel in Fig. 7). Even if the algorithm would be capable of multiple detections, the fact that the observations are defocused would make it impossible to tell how many trails are actually present and which satellite caused which trail.

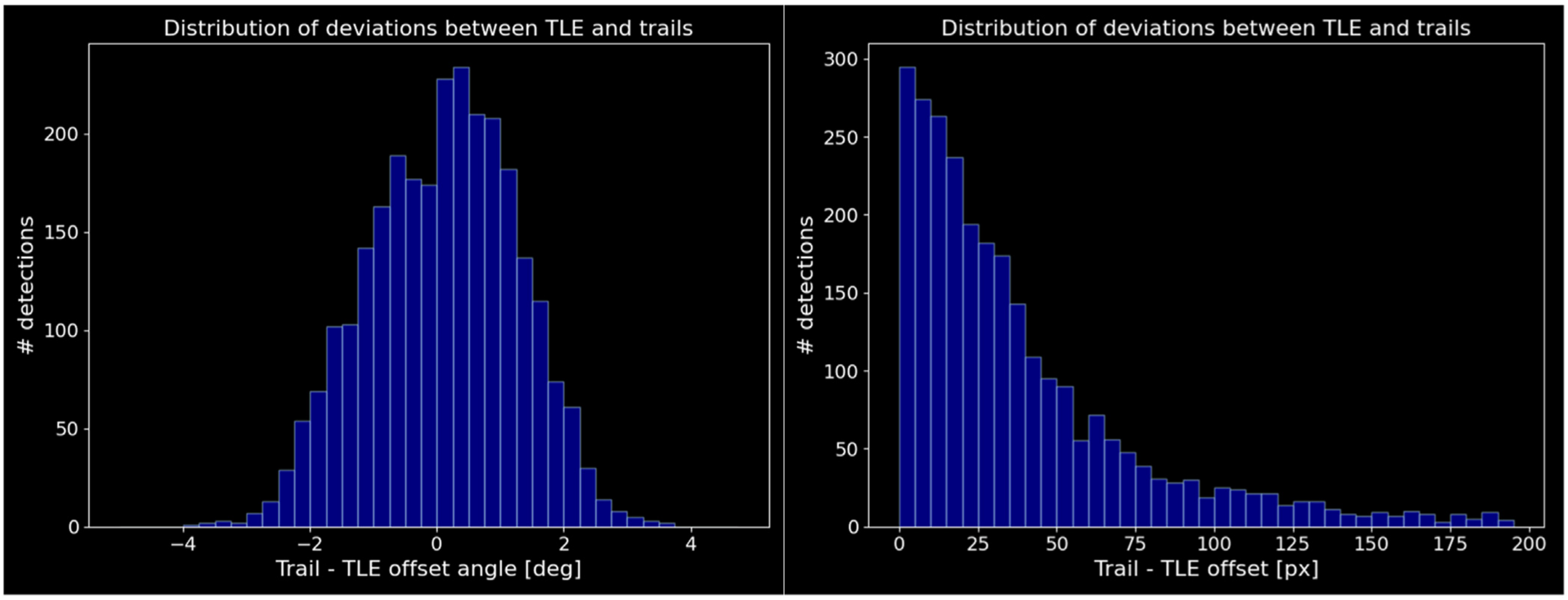

**Fig. 8.** Distributions of the offset in position angle (left) and the distance between trail and projected trajectory (right) of the identified trails (1 px = 1 arcsec).

Fig. 7 shows two examples of the trail identification process. The dashed red line corresponds to the detected trail and the solid lines to cataloged objects crossing the field during the exposure. The red solid line corresponds to the cataloged object that was identified with the detected trail. In order to account for the inaccuracy of the projection, we used margins of 350 arcsecond for the offset and 4° for the difference in position angle between trail and projected object for the assignment. The margins were chosen to account for an expected uncertainty in the positions computed from propagating the TLEs of about 1 km at 600 km distance and a rotation rate of CHEOPS of 6°/minute on average. In case there were multiple known objects within these margins, the one closest to the detected trail was selected. The distribution of pixel offset (1 px = 1 arcsec) and difference of the position angles of the identified objects is shown in Fig. 8. The majority of identifications are actually much closer to the detected track than the accepted limits, suggesting that the TLEs selected for identification generally have good accuracy.

As already explained in Section 4.4, we expect that a significant number of trails were caused by satellites from the Starlink constellation. Indeed, out of the 2742 trails that could be identified with known objects, 805 were caused by Starlinks, making the Starlink constellation responsible for about one third of the detected trails. Considering only trails caused by payloads (1799), Starlink satellites are responsible for almost 50% of the detected trails. We further identified 785 trails belonging to rocket bodies and 140 to orbital debris[3].

## 6. Conclusions

Astronomical observations have been increasingly affected by resident space objects over the past few years. Even sending telescopes into low Earth orbit does not make them immune to intruding objects crossing their field-of-view. The current projection for the next decade, including all planned private and governmental mega-constellations, is that the low Earth orbit could host hundreds of thousands of satellites. This is two orders of magnitude larger than it is today. If CHEOPS were to still be in operations then, the fraction of affected images would go from a fraction of a percent to nearly a quarter of the collected images, which would seriously impair the science case of CHEOPS. In the meantime we are exploring an alternative way to extract the photometry from CHEOPS images by fitting the Point-Spread-Function [6]. This promising method shows to be more immune to the linear contaminations than the traditional aperture photometry approach.

Despite its detrimental impact on observational astronomy, serendipitous observations of satellites or space debris by astronomical facilities can be turned into useful information. The dataset presented in this article is indeed valuable to the Space Situational Awareness community because it provides *in situ* observations of satellites and space debris over a period of 3 years that covers the start of the mega-constellation era. More generally *in situ* space surveillance is complementary to ground-observations, as illustrated by the number of companies currently developing such dedicated *in situ* missions [5]. In that respect the CHEOPS experience with observations of resident space objects is timely and of interest to some actors of the space sector, e.g. the VISDOMS team [7].

The dataset presented in this article will be made public via the International Astronomical Union's CPS[4] (Centre for the Protection of the Dark and Quiet Sky from Satellite Constellation Interference) for interested parties to use.

[3] Following the classification by space track, available at https://www.space-track.org/documents/SFS_Handbook_For_Operators_V1.7.pdf.

[4] https://cps.iau.org.

## Declaration of competing interest

The authors declare that they have no known competing financial interests or personal relationships that could have appeared to influence the work reported in this paper.

## CRediT authorship contribution statement

**Nicolas Billot:** Writing – review & editing, Writing – original draft, Visualization, Validation, Supervision, Software, Methodology, Investigation, Formal analysis, Data curation, Conceptualization. **Stephan Hellmich:** Writing – review & editing, Writing – original draft, Visualization, Validation, Supervision, Software, Methodology, Investigation, Formal analysis, Conceptualization. **Willy Benz:** Writing – original draft, Resources, Funding acquisition, Conceptualization. **Andrea Fortier:** Resources. **David Ehrenreich:** Resources. **Christopher Broeg:** Resources, Conceptualization. **Alexis Heitzmann:** Resources. **Anja Bekkelien:** Resources. **Alexis Brandeker:** Resources, Data curation. **Yann Alibert:** Resources. **Roi Alonso:** Resources. **Tamas Bárczy:** Resources. **David Barrado Navascues:** Resources. **Susana C.C. Barros:** Resources. **Wolfgang Baumjohann:** Resources. **Federico Biondi:** Resources. **Luca Borsato:** Resources. **Andrew Collier Cameron:** Resources. **Carlos Corral van Damme:** Resources. **Alexandre C.M. Correia:** Resources. **Szilard Csizmadia:** Resources. **Patricio E. Cubillos:** Resources. **Melvyn B. Davies:** Resources. **Magali Deleuil:** Resources. **Adrien Deline:** Resources. **Olivier D.S. Demangeon:** Resources. **Brice-Olivier Demory:** Resources. **Aliz Derekas:** Resources. **Billy Edwards:** Resources. **Jo Ann Egger:** Resources. **Anders Erikson:** Resources. **Luca Fossati:** Resources. **Malcolm Fridlund:** Resources. **Davide Gandolfi:** Resources. **Kosmas Gazeas:** Resources. **Michaël Gillon:** Resources. **Manuel Güdel:** Resources. **Maximilian N. Günther:** Resources. **Ch. Helling:** Resources. **Kate G. Isaak:** Resources. **Laszlo L. Kiss:** Resources. **Judith Korth:** Resources. **Kristine W.F. Lam:** Resources. **Jacques Laskar:** Resources. **Alain Lecavelier des Etangs:** Resources. **Monika Lendl:** Resources. **Demetrio Magrin:** Resources. **Pierre F.L. Maxted:** Resources. **Marko Mecina:** Resources. **Bruno Merín:** Resources. **Christoph Mordasini:** Resources. **Valerio Nascimbeni:** Resources. **Göran Olofsson:** Resources. **Roland Ottensamer:** Resources. **Isabella Pagano:** Resources. **Enric Pallé:** Resources. **Gisbert Peter:** Resources. **Daniele Piazza:** Resources. **Giampaolo Piotto:** Resources. **Don Pollacco:** Resources. **Didier Queloz:** Resources. **Roberto Ragazzoni:** Resources. **Nicola Rando:** Resources. **Heike Rauer:** Resources. **Ignasi Ribas:** Resources. **Martin Rieder:** Resources. **Nuno C. Santos:** Resources. **Gaetano Scandariato:** Resources. **Damien Ségransan:** Resources. **Attila E. Simon:** Resources. **Alexis M.S. Smith:** Resources. **Sérgio G. Sousa:** Resources. **Manu Stalport:** Resources. **Sophia Sulis:** Resources. **Gyula M. Szabó:** Resources. **Stéphane Udry:** Resources. **Bernd Ulmer:** Resources. **Solène Ulmer-Moll:** Resources. **Valérie Van Grootel:** Resources. **Julia Venturini:** Resources. **Eva Villaver:** Resources. **Nicholas A. Walton:** Resources. **Thomas G. Wilson:** Resources.

## Acknowledgment

CHEOPS is an ESA mission in partnership with Switzerland with important contributions to the payload and the ground segment from Austria, Belgium, France, Germany, Hungary, Italy, Portugal, Spain, Sweden, and the United Kingdom. The CHEOPS Consortium would like to gratefully acknowledge the support received by all the agencies, offices, universities, and industries involved. Their flexibility and willingness to explore new approaches were essential to the success of this mission. CHEOPS data analysed in this article will be made available in the CHEOPS mission archive (https://cheops.unige.ch/archive_browser/).

This project has received funding from the Swiss National Science Foundation for project 200021_200726. It has also been carried out within the framework of the National Centre of Competence in Research PlanetS supported by the Swiss National Science Foundation under grant 51NF40_205606. The authors acknowledge the financial support of the SNSF.

CBr and ASi acknowledge support from the Swiss Space Office through the ESA PRODEX program.

ABr was supported by the SNSA.

YAl acknowledges support from the Swiss National Science Foundation (SNSF) under grant 200020_192038.

We acknowledge financial support from the Agencia Estatal de Investigación of the Ministerio de Ciencia e Innovación MCIN/AEI/10.13039/501100011033 and the ERDF "A way of making Europe" through projects PID2019-107061GB-C61, PID2019-107061GB-C66, PID2021-125627OB-C31, and PID2021-125627OB-C32, from the Centre of Excellence "Severo Ochoa" award to the Instituto de Astrofísica de Canarias (CEX2019-000920-S), from the Centre of Excellence "María de Maeztu" award to the Institut de Ciències de l'Espai (CEX2020-001058-M), and from the Generalitat de Catalunya/CERCA programme.

DBa, EPa, and IRi acknowledge financial support from the Agencia Estatal de Investigación of the Ministerio de Ciencia e Innovación MCIN/AEI/10.13039/501100011033 and the ERDF "A way of making Europe" through projects PID2019-107061GB-C61, PID2019-107061GB-C66, PID2021-125627OB-C31, and PID2021-125627OB-C32, from the Centre of Excellence "Severo Ochoa" award to the Instituto de Astrofísica de Canarias (CEX2019-000920-S), from the Centre of Excellence "María de Maeztu" award to the Institut de Ciències de l'Espai (CEX2020-001058-M), and from the Generalitat de Catalunya/CERCA programme.

S.C.C.B. acknowledges support from FCT through FCT contracts nr. IF/01312/2014/CP1215/CT0004.

LBo, VNa, IPa, GPi, RRa, and GSc acknowledge support from CHEOPS ASI-INAF agreement n. 2019-29-HH.0.

ACC acknowledges support from STFC consolidated grant number ST/V000861/1, and UKSA grant number ST/X002217/1.

ACMC acknowledges support from the FCT, Portugal, through the CFisUC projects UIDB/04564/2020 and UIDP/04564/2020, with DOI identifiers 10.54499/UIDB/04564/2020 and 10.54499/UIDP/04564/2020, respectively.

A.C., A.D., B.E., K.G., and J.K. acknowledge their role as ESA-appointed CHEOPS Science Team Members.

P.E.C. is funded by the Austrian Science Fund (FWF) Erwin Schroedinger Fellowship, program J4595-N.

This project was supported by the CNES.

This work was supported by FCT - Fundação para a Ciência e a Tecnologia through national funds and by FEDER through COMPETE2020 through the research grants UIDB/04434/2020, UIDP/04434/2020, 2022.06962.PTDC.

O.D.S.D. is supported in the form of work contract (DL 57/2016/CP1364/CT0004) funded by national funds through FCT.

B.-O. D. acknowledges support from the Swiss State Secretariat for Education, Research and Innovation (SERI) under contract number MB22.00046.

ADe, BEd, KGa, and JKo acknowledge their role as ESA-appointed CHEOPS Science Team Members.

MF and CMP gratefully acknowledge the support of the Swedish National Space Agency (DNR 65/19, 174/18).

DG gratefully acknowledges financial support from the CRT foundation under Grant No. 2018.2323 "Gaseousor rocky? Unveiling the nature of small worlds".

M.G. is an F.R.S.-FNRS Senior Research Associate.

MNG is the ESA CHEOPS Project Scientist and Mission Representative, and as such also responsible for the Guest Observers (GO) Programme. MNG does not relay proprietary information between the GO and Guaranteed Time Observation (GTO) Programmes, and does not decide on the definition and target selection of the GTO Programme.

CHe acknowledges support from the European Union H2020-MSCA-ITN-2019 under Grant Agreement no. 860470 (CHAMELEON).

KGI is the ESA CHEOPS Project Scientist and is responsible for the ESA CHEOPS Guest Observers Programme. She does not participate in, or contribute to, the definition of the Guaranteed Time Programme of the CHEOPS mission through which observations described in this paper have been taken, nor to any aspect of target selection for the programme.

K.W.F.L. was supported by Deutsche Forschungsgemeinschaft grants RA714/14-1 within the DFG Schwerpunkt SPP 1992, Exploring the Diversity of Extrasolar Planets.

This work was granted access to the HPC resources of MesoPSL financed by the Region Ile de France and the project Equip@Meso (reference ANR-10-EQPX-29-01) of the programme Investissements d'Avenir supervised by the Agence Nationale pour la Recherche.

ML acknowledges support of the Swiss National Science Foundation under grant number PCEFP2_194576.

PM acknowledges support from STFC research grant number ST/R000638/1.

This work was also partially supported by a grant from the Simons Foundation (PI Queloz, grant number 327127).

NCSa acknowledges funding by the European Union (ERC, FIERCE, 101052347). Views and opinions expressed are however those of the author(s) only and do not necessarily reflect those of the European Union or the European Research Council. Neither the European Union nor the granting authority can be held responsible for them.

S.G.S. acknowledge support from FCT through FCT contract nr. CEECIND/00826/2018 and POPH/FSE (EC).

The Portuguese team thanks the Portuguese Space Agency for the provision of financial support in the framework of the PRODEX Programme of the European Space Agency (ESA) under contract number 4000142255.

GyMSz acknowledges the support of the Hungarian National Research, Development and Innovation Office (NKFIH) grant K-125015, a a PRODEX Experiment Agreement No. 4000137122, the Lendulet LP2018-7/2021 grant of the Hungarian Academy of Science and the support of the city of Szombathely.

V.V.G. is an F.R.S-FNRS Research Associate.

JV acknowledges support from the Swiss National Science Foundation (SNSF) under grant PZ00P2_208945.

NAW acknowledges UKSA grant ST/R004838/1.

TWi acknowledges support from the UKSA and the University of Warwick.